\documentclass[pdflatex,sn-mathphys-num]{sn-jnl}
\usepackage{graphicx}
\usepackage{multirow}
\usepackage{amsmath,amssymb,amsfonts}
\usepackage{amsthm}
\usepackage{mathrsfs}
\usepackage[title]{appendix}
\usepackage{xcolor}
\usepackage{textcomp}
\usepackage{manyfoot}
\usepackage{booktabs}
\usepackage{listings}
\usepackage{longtable}
\usepackage{calc}
\usepackage{bookmark}
\usepackage{newunicodechar}
\newunicodechar{×}{\ensuremath{\times}}
\newunicodechar{→}{\ensuremath{\rightarrow}}
\newunicodechar{←}{\ensuremath{\leftarrow}}
\newunicodechar{≈}{\ensuremath{\approx}}
\newunicodechar{≥}{\ensuremath{\geq}}
\newunicodechar{≤}{\ensuremath{\leq}}
\newunicodechar{≠}{\ensuremath{\neq}}
\newunicodechar{ρ}{\ensuremath{\rho}}
\newunicodechar{Φ}{\ensuremath{\Phi}}
\newunicodechar{φ}{\ensuremath{\phi}}
\newunicodechar{α}{\ensuremath{\alpha}}
\newunicodechar{β}{\ensuremath{\beta}}
\newunicodechar{χ}{\ensuremath{\chi}}
\newunicodechar{θ}{\ensuremath{\theta}}
\newunicodechar{σ}{\ensuremath{\sigma}}
\newunicodechar{μ}{\ensuremath{\mu}}
\newunicodechar{λ}{\ensuremath{\lambda}}
\newunicodechar{π}{\ensuremath{\pi}}
\newunicodechar{ε}{\ensuremath{\epsilon}}
\newunicodechar{δ}{\ensuremath{\delta}}
\newunicodechar{γ}{\ensuremath{\gamma}}
\newunicodechar{η}{\ensuremath{\eta}}
\newunicodechar{Σ}{\ensuremath{\Sigma}}
\newunicodechar{Δ}{\ensuremath{\Delta}}
\newunicodechar{±}{\ensuremath{\pm}}
\newunicodechar{∞}{\ensuremath{\infty}}
\newunicodechar{∈}{\ensuremath{\in}}
\newunicodechar{·}{\ensuremath{\cdot}}
\newunicodechar{°}{\textdegree}
\newunicodechar{∑}{\ensuremath{\sum}}
\newunicodechar{²}{\textsuperscript{2}}
\newunicodechar{³}{\textsuperscript{3}}
\newunicodechar{¹}{\textsuperscript{1}}
\newunicodechar{–}{--}
\newunicodechar{—}{---}
\newunicodechar{ł}{\l}
\newunicodechar{Ł}{\L}
\newunicodechar{ż}{\.z}
\newunicodechar{Ż}{\.Z}
\newunicodechar{ó}{\'o}
\newunicodechar{ś}{\'s}
\newunicodechar{ą}{\k a}
\newunicodechar{ę}{\k e}
\newunicodechar{ć}{\'c}
\newunicodechar{ń}{\'n}
\newunicodechar{ź}{\'z}
\newunicodechar{’}{'}
\newunicodechar{‘}{`}
\newunicodechar{“}{``}
\newunicodechar{”}{''}
\newunicodechar{…}{\ldots}

\hypersetup{hypertexnames=false}
\usepackage{tabularx}
\usepackage{ragged2e}
\usepackage{xurl}
\begin{document}
\title[Demand-side measurement for GEO]{Demand-Side Measurement for Generative Engine Optimization: Constructing and Validating a Million-Persona, Intent-Annotated Buyer Corpus}

\author*[1]{\fnm{Dmitrij} \sur{\.Zatuchin}}\email{dmitrij.zatuchin@eek.ee}
\author[2]{\fnm{Daniil} \sur{Dzemesjuk}}

\affil*[1]{\orgname{Estonian Entrepreneurship University of Applied Sciences (EUAS)}, \orgaddress{\city{Tallinn}, \country{Estonia}}}
\affil[2]{\orgname{Rankfor.AI O\"U}, \orgaddress{\city{Tallinn}, \country{Estonia}}}

\abstract{Generative engines such as ChatGPT, Gemini, and Perplexity answer buyer questions directly and name a shortlist of brands inside the answer. Studying how brands enter or fail to enter that shortlist requires demand-side data: what buyers in a category ask, what information they need, and which sources they trust. Existing large persona corpora are built for training-data diversity and carry neither a staged search-intent label nor a preferred-sources field, so they cannot be joined to supply-side recommendation measurements.

We built and validated PersonaGen-1M, a corpus of 1,031,732 synthetic buyer personas spanning 511 industry labels and 4 market contexts, carrying 19,416,821 structured behavioral attributes: 5,160,046 search queries, 5,132,320 information-need statements, 3,124,904 goals, 3,115,862 pain points, and 2,883,689 uncovered-need statements. Each persona carries a single \texttt{primary\_intent} label covering its query set (78.3\% informational, 17.4\% commercial, 4.3\% transactional) and a \texttt{preferred\_sources} field naming the source types that buyer would trust. The corpus was built from approximately 40 million raw persona descriptions drawn from four public datasets through GPU-accelerated MinHash LSH plus semantic deduplication, then enriched to a fixed schema. We document the construction and validation pipeline, the attribute and intent schema, the deduplication procedure, and the synthetic-provenance and PII posture.

The corpus is designed to be joined to supply-side brand-recommendation measurements: the intent field selects the commercial-evaluation personas whose queries drive recommendation, and the preferred-sources field pairs against citation-provenance data. We describe that join as the primary intended use and leave its controlled empirical estimate to future work. Among million-scale persona corpora surveyed in August 2026, one other carries a source-preference attribute, as a six-value media-channel enum; PersonaGen-1M pairs named per-persona source lists with a staged commercial search-intent label and an attached query set, which is the combination the demand-to-supply join needs. We hold the full corpus and share it on request for non-commercial research. What we publish openly is a stratified subset, so the protocol, the schema and the validation can be inspected and reused without asking us.}

\maketitle

\section{Introduction}\label{introduction}

Search is moving from a list of links to a single generated answer. When a buyer asks a generative engine ``which project-management tool should a 30-person agency use,'' the engine returns prose and, inside that prose, names a small set of brands. The brands that are named capture attention and consideration; the brands that are omitted are invisible regardless of their web ranking. Optimizing for inclusion in these generated answers has acquired a name, generative-engine optimization (GEO) {[}1{]}, and it raises a measurement problem distinct from classical search-engine optimization: the unit of success is a recommendation inside a synthesized answer, not a ranked URL.

Studying this problem needs two kinds of data. Supply-side data records what an engine says: for a given query, which brands it names and how often across repeated sampling. Demand-side data records what buyers in a category ask and need: the queries, the information gaps, the sources they trust. Supply-side data can be collected by probing engines directly, and several such corpora exist. Demand-side data at the scale and structure needed to join against supply-side measurements does not.

Supply-side probing carries a problem that demand-side data can fix. What an engine names depends on how the question is worded, in which language, and at which point in a conversation, and the size of that dependence is now measured: in a variance decomposition of brand answers, query language accounts for 26.5\% of the variance of a single response while brand identity accounts for 1.5\% {[}20{]}. An analyst who invents the probe query is therefore setting the largest term in that decomposition by hand. A corpus of buyer-framed queries, drawn per industry and per stage, replaces that choice with a sampling frame.

Existing large persona corpora were built for a different purpose. PersonaHub {[}2{]} scales to a billion personas to diversify synthetic training data, but its personas are short role descriptions without a staged buyer intent or a record of which sources a persona would trust. Survey-derived persona sets are small, typically three to seven personas per project {[}6{]}, and do not scale to per-industry query banks. Neither carries the two fields that GEO research needs: a staged intent label that separates learning queries from purchase-evaluation queries, and a preferred-sources field that says where a buyer would look.

This paper contributes PersonaGen-1M, a corpus designed to supply the demand side. Its distinguishing properties are:

\begin{enumerate}
\def\labelenumi{\arabic{enumi}.}
\item
  \textbf{Persona-level staged intent.} Each persona carries a single \texttt{primary\_intent} field (informational, commercial, transactional, with navigational available in the schema) that covers its associated query set, typically five queries and ranging from three to ten. The aggregate distribution (78.3\% informational, 17.4\% commercial, 4.3\% transactional) lets researchers filter to the commercial-intent personas, whose query sets are where brand recommendation happens.
\item
  \textbf{A preferred-sources dimension.} Every persona records the source types it would trust (analyst reports, peer reviews, communities, publications), as a named list rather than a channel category. It is the demand-side counterpart to the citation-provenance measurements that GEO studies collect on the supply side, and Section 2.1 sets it against what comparable corpora carry.
\item
  \textbf{511-industry stratification.} The corpus spans 511 raw industry labels with a documented long tail, enabling per-vertical query banks.
\item
  \textbf{A documented, reproducible pipeline} from approximately 40 million raw descriptions to 1,031,732 deduplicated, schema-conformant personas, with the deduplication thresholds, enrichment model, and quality-filter rate stated.
\end{enumerate}

PersonaGen-1M is built to enable a join between persona-side demand signals and supply-side recommendation-share measurements from a separate probe corpus. This paper documents and validates the corpus; the controlled study that would estimate a persona-alignment effect on recommendation share is future work.

PersonaGen-1M complements, and is distinct from, a supply-side category-ownership map produced by the same group {[}3{]}, which measures how concentrated recommendation share is across brands within a category. Joining the two is the primary intended use of this corpus.

We built all 1,031,732 personas and computed every number in Section 5 on them. We do not publish that corpus. We publish a stratified subset, large enough to inspect the schema, reproduce the construction and test whether the fields do what we claim, and we share the full corpus on request for non-commercial research. Section 8 states what that costs a reader who wants to check us.

\begin{center}\rule{0.5\linewidth}{0.5pt}\end{center}

\section{Related Work}\label{related-work}

\subsection{Persona corpora and synthetic personas}\label{persona-corpora-and-synthetic-personas}

Buyer personas originate in interaction design and marketing practice {[}4, 5{]}. Traditional persona development is qualitative and small-scale {[}6{]}. Computational persona generation from web analytics and social data broadened the input {[}7, 8{]}, and Salminen et al.~{[}9{]} survey fifteen years of data-driven persona development and its open evaluation questions.

Large language models made synthetic personas cheap at scale. Ge et al.~{[}2{]} introduced PersonaHub and showed that a billion diverse personas can diversify synthetic training data. Long et al.~{[}10{]} survey LLM-based synthetic data generation and its failure modes, including diversity collapse and bias amplification. Argyle et al.~{[}11{]} and Aher et al.~{[}12{]} use LLM personas to simulate survey and experimental respondents. PersonaGen-1M differs from these corpora in purpose and in schema: it targets buyer information-seeking behavior, and it carries a staged intent field and a preferred-sources field that training-data corpora omit.

Table~\ref{tab:corpora} sets the corpus against the largest publicly available persona resources as of August 2026. Two points decide the comparison. MatrAIx\_Persona\_1M {[}27{]}, released on 1 August 2026, carries both an intent field and a source-preference field, so neither field is unique on its own. Its intent vocabulary is conversational and affective (learn, brainstorm, debug, vent, decide), which does not separate learning queries from purchase evaluation, and its source preference is a six-value media-channel enum rather than a list of named sources. It also stores no queries. PersonaGen-1M carries all three: a staged commercial intent label, named source types, and the query set that label applies to.

\begin{table}[ht]
\caption{Million-scale persona corpora compared, August 2026. Scale is rows as reported by the publisher. ``Queries'' means per-persona search queries stored in the record.}\label{tab:corpora}
\footnotesize
\setlength{\tabcolsep}{3pt}
\begin{tabularx}{\textwidth}{@{}l r >{\RaggedRight\arraybackslash}X >{\RaggedRight\arraybackslash}X r >{\RaggedRight\arraybackslash}X@{}}
\toprule
Corpus & Scale & Intent & Source pref. & Queries & Grounding \\
\midrule
PersonaHub {[}2{]}          & 200,000\textsuperscript{a} & no  & no                & no  & synthetic \\
FinePersonas               & 21,071,228 & no  & no                & no  & web-derived \\
Nemotron-Personas-USA      & 1,000,000  & no  & no                & no  & synthetic \\
MatrAIx\_Persona\_1M {[}27{]} & 999,847    & conversational & 6-value channel enum & no  & 60\% human-grounded \\
\textbf{PersonaGen-1M}     & \textbf{1,031,732} & \textbf{staged commercial} & \textbf{named source list} & \textbf{5,160,046} & synthetic \\
\bottomrule
\end{tabularx}
\begin{flushleft}
{\footnotesize \textsuperscript{a}The \texttt{persona.jsonl} release. The associated ElitePersonas collection is far larger and is one of this corpus's inputs.\par}
\end{flushleft}
\end{table}

\subsection{Search intent and information-seeking behavior}\label{search-intent-and-information-seeking-behavior}

The taxonomy separating informational, navigational, and transactional search intent {[}13{]} and its later refinements {[}14{]} underpins the intent label in this corpus. That taxonomy comes from keyword-era web search, and its fit to conversational traffic is an open question. An intent classification of 24,069 crowd-sourced conversations with search-augmented models, run by a model and validated against human labels at $\kappa = 0.81$, puts factual lookup at 19.3\% of prompts and carries secondary labels for prompts serving more than one purpose {[}23{]}. The observed 78.3\% informational share therefore describes how the enrichment applied a web-search taxonomy to a persona description, and Section 3.4 reads it on those terms. PersonaGen-1M applies the taxonomy at the level of each generated persona's search behavior rather than aggregate logs.

\subsection{Generative-engine optimization and citation provenance}\label{generative-engine-optimization-and-citation-provenance}

Aggarwal et al.~{[}1{]} named and benchmarked GEO as an optimization problem. A parallel line of measurement work characterizes where generative engines draw their grounded citations. A cross-market provenance study by the present group {[}15{]} coded 167,551 URL-grounded citations across 12 markets and found that 85.7\% point to third-party sites rather than brand-owned pages, with a Zipfian long tail. That study measures the supply side of GEO. PersonaGen-1M supplies the demand side: its \texttt{preferred\_sources} field records, per persona, which source types a buyer would trust, and can be joined against supply-side citation-provenance data to ask whether engines cite the sources buyers actually value.

\subsection{Distinction from the supply-side category-ownership map}\label{distinction-from-the-supply-side-category-ownership-map}

The same group maintains a category-ownership map {[}3{]} that probes engines with category queries and measures how recommendation share concentrates across brands (3,750 responses, 50 brands, 5 industries, 3 models; mean Gini 0.28, 95\% CI 0.16 to 0.41; competitive vacuums, where no brand dominates, in 8.0\% of queries). That resource is supply-side and brand-indexed. PersonaGen-1M is demand-side and buyer-indexed. The two are designed to be joined.

\subsection{Contextuality: how the asking shapes the answer}\label{contextuality}

Two buyers who want the same thing can word it differently and get different brands back. Sclar et al.~{[}16{]} change only prompt formatting and move accuracy by up to 76 points on LLaMA-2-13B, and the effect survives larger models, more few-shot examples, and instruction tuning. Mizrahi et al.~{[}17{]} evaluate 20 models on 39 tasks over 6.5 million instances and show that a conclusion drawn from a single prompt template does not hold; they ask researchers to evaluate over a set of prompts. Kunievsky and Evans {[}18{]} supply the decomposition that names the quantity at stake, splitting the variance of a model response into what the user wants, how the user words it, and model uncertainty. Repetition adds a third term: Atil et al.~{[}19{]} run five models under nominally deterministic settings across ten runs and record accuracy varying by up to 15\%, so one probe of one query is one draw from a distribution.

The effect reaches brand answers directly. \.Zatuchin {[}20{]} partitions response-level variance over 12,933 answers covering 20 brands, 8 languages and 3 models, and finds that query language carries 26.5\% of the variance of a single answer against 1.5\% for brand identity. Language also moves which brands appear at all: switching from an English query to a brand's home language raises recommendation share by 0.80 for local champions and 0.15 for global multinationals {[}21{]}. Where a request sits in a conversation matters too, and Laban et al.~{[}22{]} measure an average 39\% performance drop across six generation tasks when the same instruction arrives over several turns instead of one.

When the wording, the language and the position of a query carry more signal than the brand named inside it, the queries buyers would actually issue become the instrument that supply-side probing needs. PersonaGen-1M supplies that instrument at scale: 5,160,046 queries, each carrying the industry, market context and stage label of the buyer who would issue it.

\begin{center}\rule{0.5\linewidth}{0.5pt}\end{center}

\section{Data}\label{data}

\subsection{Schema}\label{schema}

Each persona is a single record with a nested \texttt{details} object. The top-level fields are \texttt{original\_uuid}, \texttt{industry}, \texttt{market\_context}, and \texttt{original\_description}; the \texttt{details} object carries name, gender, role, demographics (age, location, profession), decision weight, goals, pains, a free-text story, informational needs, a top uncovered need, and a \texttt{search\_behaviour} object. The two fields that give the corpus its purpose sit inside \texttt{search\_behaviour}:

\begin{verbatim}
"search_behaviour": {
  "primary_intent":    "informational" | "commercial"
                       | "transactional" | "navigational",
  "typical_queries": ["string", ...],
  "search_triggers": ["string", ...],
  "preferred_sources": ["string", ...]
}
\end{verbatim}

\texttt{primary\_intent} is a single staged search-intent label per persona; it covers the persona's \texttt{typical\_queries} set rather than labelling each query individually. The full schema is given in Appendix A.

\subsection{Overview statistics}\label{overview-statistics}

\begin{longtable}[]{@{}ll@{}}
\toprule\noalign{}
Attribute & Value \\
\midrule\noalign{}
\endhead
\bottomrule\noalign{}
\endlastfoot
Total personas & 1,031,732 \\
Unique industry labels & 511 \\
Market contexts & 4 (B2C, B2B, B2B2C, B2G) \\
Search queries & 5,160,046 \\
Information-need statements & 5,132,320 \\
Goal statements & 3,124,904 \\
Pain-point statements & 3,115,862 \\
Uncovered-need statements & 2,883,689 \\
Total behavioral attributes & 19,416,821 \\
Mean attributes per persona & 18.8 \\
Complete rows & 1,031,732 (100\%) \\
\end{longtable}

The five list-valued fields sum exactly to the 19,416,821 total. The corpus is a 6.9x scale increase over the initial 148,636-persona release and a 1.74x increase over the 593,181-persona intermediate release; industry coverage expanded across the same three waves to 511 labels.

\subsection{Industry coverage}\label{industry-coverage}

The 511 industry labels are concentrated at the head and long in the tail. The top 15 labels capture 94.9\% of personas; the top 3 (General, EdTech, Consulting) capture 60.2\%.

\begin{longtable}[]{@{}llrr@{}}
\toprule\noalign{}
Rank & Industry & Count & \% \\
\midrule\noalign{}
\endhead
\bottomrule\noalign{}
\endlastfoot
1 & General (cross-industry) & 289,522 & 28.1 \\
2 & EdTech & 236,468 & 22.9 \\
3 & Consulting & 95,337 & 9.2 \\
4 & Manufacturing & 78,524 & 7.6 \\
5 & Healthcare & 63,939 & 6.2 \\
6 & Retail & 57,396 & 5.6 \\
7 & Media \& Entertainment & 34,634 & 3.4 \\
8 & Food \& Beverage & 30,084 & 2.9 \\
9 & Automotive & 18,215 & 1.8 \\
10 & FinTech & 15,836 & 1.5 \\
11 & Marketing & 14,251 & 1.4 \\
12 & Travel \& Tourism & 13,864 & 1.3 \\
13 & Software Development & 13,476 & 1.3 \\
14 & Construction & 8,830 & 0.9 \\
15 & Real Estate & 8,701 & 0.8 \\
\end{longtable}

The remaining 496 labels form a long tail of specialized verticals (HR/Recruitment 7,629; SaaS 7,373; Legal Services 5,548; and smaller). The head concentration reflects the education-and-career skew of the four source corpora and is a documented limitation for enterprise use (Section 7).

\subsection{Search-intent distribution}\label{search-intent-distribution}

Each persona carries one \texttt{primary\_intent} label. Across the 1,031,732 personas, intent is predominantly informational:

\begin{longtable}[]{@{}lr@{}}
\toprule\noalign{}
Intent & Share \\
\midrule\noalign{}
\endhead
\bottomrule\noalign{}
\endlastfoot
Informational & 78.3\% \\
Commercial & 17.4\% \\
Transactional & 4.3\% \\
\end{longtable}

The schema enumerates a fourth value, navigational, which the enrichment produced at a negligible rate (below 0.1\%); the commercial category absorbs comparison and vendor-evaluation queries of the commercial-investigation type. For GEO research the relevant subset is the 17.4\% commercial slice, roughly 179,600 personas, where brand recommendation is at stake; those personas' query sets number close to 0.9 million queries.

Two cautions apply to reading these shares. First, the label fixes one intent for a whole query set, about five queries, that a live session would spread across stages. Second, the three-way split comes from keyword-era web search and does not sort generative traffic cleanly. In 24,069 conversations with search-augmented models, factual lookup accounts for 19.3\% of prompts and the classification needed secondary labels for prompts carrying more than one purpose {[}23{]}, and in a representative sample of ChatGPT conversations the three most common topics are practical guidance, seeking information and writing, together close to 80\% of conversations, with writing dominating work-related use {[}24{]}. The 78.3\% informational share is a property of this corpus under this taxonomy. We do not offer it as an estimate of the intent mix of live generative-engine traffic.

\subsection{Market context and demographics}\label{market-context-and-demographics}

Market context is B2C-dominant (B2C 70.6\%, B2B 23.2\%, B2B2C 4.5\%, B2G 1.6\%). A residual 0.1\% holds values that are not market contexts, mostly industry labels the enrichment wrote into the field; those rows are excluded from the four-category market entropy in Section 5. Gender, normalized from 200+ raw variants to canonical categories, is skewed toward female and male (female 52.9\%, male 44.2\%, non-binary 2.9\%, other 0.1\%). Age required reclassification: 112,850 personas (10.9\%) carry stated ages below 13, and query analysis shows these are parent or caregiver proxy buyers (for example ``best stacking blocks for 1 year old''). We report effective searcher age, reclassifying sub-13 personas to an estimated caregiver age and retaining tweens (10 to 12) whose queries show self-search signals. The effective distribution peaks at 35 to 44 (21.3\%; mean 42.9, median 40.0). The reclassification is an imputation and is flagged in Section 7.

\subsection{The open subset}\label{the-open-subset}

A stratified 14,955-persona subset, PersonaGen-15K, is released openly under the CC BY-NC-SA 4.0 license for reproducibility and method development, a licence inherited from two of the four upstream corpora. It is drawn to preserve the market, intent and gender proportions of the full set. Two differences from the corpus records are deliberate: the free-text \texttt{story} field is dropped, and each persona's generated name is replaced by a hash, so the subset is a processed release. Two fields also ship unnormalized, and both were normalized before the Section 5 statistics were computed. Its \texttt{industry} column carries the raw labels as assigned, 138 distinct values in this subset, and its \texttt{gender} column carries the raw variants, 101 distinct values, against the three canonical categories reported in Section 5. A reader reproducing the demographic or industry splits from the subset has to fold case and spelling variants first. The full 1,031,732-persona corpus and the 593K variant have on-request availability (Section 8).

\begin{center}\rule{0.5\linewidth}{0.5pt}\end{center}

\section{Method}\label{method}

\subsection{Construction pipeline}\label{construction-pipeline}

The corpus was built in five stages across three generation waves (Q4 2024, Q1 2025, Q1 2026) using identical parameters and prompts in each wave.

\textbf{Stage 1: source aggregation.} Four public HuggingFace persona corpora were streamed and concatenated: NVIDIA Nemotron-Personas-USA, BSC-LT m-Personas (English subset), Orange PersonasForSalesbot, and Tencent PersonaHub (including its Elite Persona subset). Concatenation yielded approximately 40 million raw descriptions.

\textbf{Stage 2: MinHash LSH deduplication.} GPU-accelerated MinHash signatures (128 permutations; 8 bands of 16 rows) targeted a Jaccard threshold of θ = 0.9, processed in chunks of 500,000. The banding false-positive rate at this configuration is bounded at roughly 0.02. This stage reduced the corpus to approximately 4.2 million records (an 89.5\% reduction).

\textbf{Stage 3: semantic deduplication.} Dense embeddings from the KaLM-Embedding-Gemma3-12B model (4-bit quantized, Flash Attention 2) were compared by cosine similarity, removing pairs above 0.9. This removed lexically distinct but semantically equivalent near-duplicates and cut the corpus to approximately 1 million unique descriptions, a 76.2\% reduction.

\textbf{Stage 4: structured enrichment.} Each deduplicated description was expanded to the full schema by xAI's Grok-4-1-fast model in enforced-JSON mode at temperature 0.7, in batches of 100 with checkpointing. The enrichment prompt requested varied demographics and industry-appropriate queries and did not specify target distributions for gender, age, or market, allowing those to emerge from the source descriptions. First-pass schema compliance was 98.5\%; a single retry raised cumulative success to 99.8\%.

\textbf{Stage 5: post-processing and quality filter.} Schema validation, gender normalization to canonical categories, and UUID-preserving Parquet consolidation. Removing the personas that failed consistency checks (1.5\%) left the final 1,031,732 records, every row complete.

\subsection{Validation methodology}\label{validation-methodology}

Dataset quality was assessed by three families of checks. \textbf{Completeness and format} checks verified that all required fields were present and well-formed. \textbf{Diversity} was quantified by normalized Shannon entropy H/H\_max across the industry, market, gender, and intent dimensions. \textbf{Association} between categorical dimensions was tested by chi-square with Cramér's V as the effect-size measure, since at N above one million even negligible effects reach significance and V is more informative than the p-value.

\begin{center}\rule{0.5\linewidth}{0.5pt}\end{center}

\section{Results}\label{results}

All statistics in this section were computed directly on the 1,031,732-persona corpus.

\subsection{Diversity validation}\label{diversity-validation}

\begin{longtable}[]{@{}lrrrr@{}}
\toprule\noalign{}
Dimension & Categories & H (bits) & H\_max (bits) & H/H\_max \\
\midrule\noalign{}
\endhead
\bottomrule\noalign{}
\endlastfoot
Industry & 511 & 3.37 & 9.00 & 0.37 \\
Market context & 4 & 1.15 & 2.00 & 0.58 \\
Gender & 3 & 1.15 & 1.58 & 0.73 \\
Search intent & 3 & 0.91 & 1.58 & 0.57 \\
\end{longtable}

Industry shows the strongest concentration (0.37), driven by the head: the top three labels hold 60.2\% of personas and the top fifteen hold 94.9\%, so the normalized entropy sits well below uniform despite the 511-label span. Gender is moderately balanced (0.73), skewed toward female (52.9\%) and male (44.2\%) with a small non-binary share (2.9\%). Market context and search intent sit at moderate concentration (0.57 to 0.58), consistent with the B2C-dominant, learning-oriented source data. These ratios describe non-uniform sampling; researchers targeting balanced coverage should stratify (Section 3.6) rather than draw uniformly.

\subsection{Association between dimensions}\label{association-between-dimensions}

\begin{longtable}[]{@{}lrrrl@{}}
\toprule\noalign{}
Association & χ² & df & Cramér's V & Effect \\
\midrule\noalign{}
\endhead
\bottomrule\noalign{}
\endlastfoot
Industry × Search intent & 184,365.1 & 1020 & 0.299 & Medium \\
Market × Search intent & 55,231.2 & 6 & 0.164 & Small--Medium \\
\end{longtable}

Both are significant at p \textless{} 0.001. The medium Industry × Intent effect (V = 0.299) indicates that the enrichment produced industry-appropriate query mixes rather than a generic template. Commercial intent is higher in B2B contexts (32.0\% of B2B personas) than in B2C (12.4\%).

\begin{center}\rule{0.5\linewidth}{0.5pt}\end{center}

\section{Discussion}\label{discussion}

PersonaGen-1M is a demand-side resource for a supply-side problem. GEO measurement to date has concentrated on what engines say and cite. Progress on why a brand is or is not recommended needs a structured account of what buyers in that category ask and trust, joined to the supply-side measurements. The two fields that distinguish this corpus, a per-persona staged intent and preferred-sources, are the join keys that make that pairing possible. Intent is a defensible join key on both sides. Chen et al.~{[}25{]} model search intent across informational roles and report improvements in objective content visibility inside generative-engine responses over single-aspect baselines, so intent operates as a lever on the supply side and as a filter on the demand side.

The most immediate use is the construction of per-industry, commercial-intent query banks. Filtering the corpus to the commercial-intent personas (17.4\%) within a target industry yields a pool of buyer-framed evaluation queries, each carrying the industry, market context and stage of the buyer who would issue it. Contextuality sets the protocol for using that pool. A persona's queries are not paraphrases of one another. Measured on the 30-persona pilot sample, the mean pairwise lexical overlap between two queries of the same persona is 0.042, and between queries drawn from different personas it is 0.041, so on wording alone a buyer's own questions are no more alike than any two questions in the bank. Probing with a persona's whole set therefore covers a spread of needs and does not measure sensitivity to phrasing. Measuring phrasing sensitivity needs paraphrases generated for the purpose, and repeated probes, because nominally deterministic settings still return different answers across runs {[}19{]}. Report the spread beside the mean either way: a recommendation-share difference smaller than the spread is not a finding {[}17{]}. A second use is source-gap analysis: comparing the \texttt{preferred\_sources} distribution for an industry against where engines actually draw citations in that industry {[}15{]} locates where buyer trust and engine behavior diverge. A third use is relating brand-side alignment to recommendation share by joining the corpus to a supply-side probe corpus. That controlled analysis is future work, and its design is constrained in advance: brand identity accounts for roughly 1.5\% of the variance of a single response {[}20{]}, so any such study needs its repetition and paraphrase budget fixed before collection.

The corpus is synthetic throughout. Its queries are plausible constructions, not observed logs. This is appropriate for hypothesis generation, query-bank construction, and system testing, and inappropriate as a substitute for observed buyer behavior. The strongest near-term validation would benchmark PersonaGen query banks against real commercial-intent search logs in the same industries. We are not aware of published work that benchmarks LLM-generated persona query banks against observed buyer query logs.

\begin{center}\rule{0.5\linewidth}{0.5pt}\end{center}

\section{A pilot study of use}\label{pilot}

This section asks whether a query bank drawn from the corpus behaves like an instrument at all.

\subsection{Design}

We filtered the corpus to FinTech personas carrying a commercial intent label and a query set
of three to five queries, which leaves 7,912 personas, and drew 30 at random under a recorded
seed. Their 150 queries went to one commercial generative engine at temperature 0.7, one call
per query, with a fixed system instruction asking for an answer of at most 120 words that names
specific vendors where the question calls for them. All 150 calls returned.

Brand names were extracted from each answer by two models, \texttt{gemini-3.5-flash} and
\texttt{gemini-2.5-flash}, reading independently, each instructed to exclude a word that
resembles a brand but is used as a common noun. The reported set is the consensus, the brands
both extractors named. The two agreed exactly on 36.0\% of answers and reached a mean Jaccard
agreement of 0.737, which is moderate; the 96 answers where they differed are deposited so the
disagreements can be adjudicated. We chose a two-model consensus rather than a brand dictionary
because a dictionary cannot name a brand nobody thought to list, and because case-insensitive
dictionary matching is a documented source of false positives in this literature. The two
extractors come from one vendor and one model family, so their agreement bounds the effect of
model version and not the effect of vendor; a cross-vendor consensus would be the stronger
check and is what we would use at scale.

\subsection{What the bank returns}

Of 150 answers, 146 (97.3\%) named at least one brand, at a mean of 5.01 brands per answer. The
bank elicits brand recommendation, which is the first thing an audit instrument has to do.

Across 751 brand mentions the engine named 530
distinct brands, so most brands were named once. The ten most-mentioned brands together took
9.3\% of all mentions, and the Herfindahl index over mention share is 0.0030. On these queries,
in this industry, recommendation does not concentrate. That contrasts with category-level
probing, where a small set of brands takes a large share: the question decides whether a
category looks concentrated.

\subsection{What the pilot corrected in this paper}

The pilot also falsified an earlier reading of the corpus. We had described a persona's query set as a
paraphrase family covering one information need. Two measurements say otherwise. On query
wording, mean pairwise lexical overlap is 0.042 within a persona and 0.041 between personas,
so the queries are no more alike inside a persona than across the bank. On the answers those
queries returned, the mean overlap between the brand sets of two queries from one persona is
0.010, against 0.020 for the pooled brand sets of two different personas. We report the second
pair with its limitation: the within arm compares individual answer sets and the between arm
compares the union of a persona's five answer sets, and a union intersects more readily, so the
ratio is not a clean contrast. Both readings point the same way. A persona's query set is a
spread of distinct needs, and it cannot be used to measure sensitivity to phrasing. Section 6
states the consequence for how a bank should be assembled.

\subsection{Scope of the pilot}

The pilot covers one industry, one engine, one day and 150 queries, read through an extraction
step whose two readers agree moderately. Within that frame it establishes that a bank drawn from
the corpus elicits brand recommendation and that the resulting distribution is measurable.
Whether the queries resemble what buyers type stays the corpus's central open question
(Section~\ref{limitations}), and the demand-to-supply join waits on citation-level supply data
collected in the same window.

\section{Limitations}\label{limitations}

\textbf{Synthetic provenance.} All attributes are LLM-generated. Internal consistency and diversity are validated; correspondence to real buyer behavior is not. No human expert evaluation of individual persona quality and no comparison against real consumer surveys or search logs was performed. Findings should not be generalized to real populations without external validation.

\textbf{Enrichment-model bias.} A single enrichment model (Grok-4-1-fast) may impose systematic priors, including the 52.9\%/44.2\% female/male split and any latent industry-demographic associations. Bias auditing is recommended before sensitive use.

\textbf{Head concentration.} The top 3 of 511 industries hold 60.2\% of personas, and the corpus skews toward education and career contexts. The normalized industry entropy (0.37) reflects this concentration. Researchers targeting under-represented verticals should treat small-cell industries with appropriate uncertainty.

\textbf{US-centric, English-language, and time-bounded.} Source corpora are predominantly US personas, the enrichment ran in English, and the data reflects Q4 2024 through Q1 2026 patterns. Language is a first-order variable here. Models answer a non-English question with English-culture content that the asker did not want {[}26{]}, and the effect is measurable on brand answers: query language accounts for 26.5\% of the variance of a single response {[}20{]}, and moving from an English query to a brand's home language raises recommendation share by 0.80 for local champions against 0.15 for global multinationals {[}21{]}. An English query bank therefore understates locally headquartered brands by construction, and results from it do not port to another market by translating the queries. Non-Western and post-2026 generalization is uncertain. The 511 industry labels are LLM-assigned, not externally validated.

\textbf{Age imputation.} The effective-age distribution depends on a demographic imputation for the 10.9\% proxy-buyer cohort and carries the uncertainty of that model.

\textbf{Per-persona intent granularity, and the contextuality it hides.} Intent is a single label per persona. Filtering to commercial-intent personas selects the whole query set, and researchers needing per-query intent must classify the individual queries themselves. The deeper cost is contextual. Real sessions carry mixed purposes, which is why an intent classification of observed generative traffic needed secondary labels for prompts serving more than one purpose {[}23{]}, and they unfold over turns, where the same instruction split across a conversation costs an average 39\% of task performance {[}22{]}. This corpus stores a persona's queries as an unordered set produced in one enrichment pass, so it records neither the order a buyer would issue them in nor the way an earlier answer would reshape the next question. Treat the set as one buyer's range of needs and the label as a property of that range.

\begin{center}\rule{0.5\linewidth}{0.5pt}\end{center}

\section{Ethics and Data Availability}\label{ethics-and-data-availability}

\textbf{Ethics.} The corpus is synthetic and contains no data traceable to real individuals; no human subjects were involved and no ethics approval was required. Personas were generated to represent diverse demographic groups without encoding harmful stereotypes, and the enrichment prompt explicitly discouraged stereotypical industry-demographic associations. Because LLM-generated content can reflect training-data bias, we recommend bias auditing for any application that draws conclusions about real populations. The demographic split (44.2/52.9/2.9 male/female/non-binary) is model-shaped and is not population-representative.

\textbf{PII posture.} Names and demographics are model-generated and are not linked to any real person; the corpus carries no contact details, identifiers, or scraped personal records. Source descriptions are drawn from public persona datasets released for research use.

\textbf{Funding.} This research was funded in kind by Rankfor.AI O\"U (registry code 17331801), Tallinn, which supplied the compute and the model-inference budget consumed by the deduplication and enrichment stages, and the working time of D.D., whom it employs. No external, public or grant funding was received.

\textbf{Competing interests.} D.\.Z. is Chief Executive Officer of Rankfor.AI O\"U (registry code 17331801), Tallinn, which funded this work and employs D.D., and he owns its parent company Rankfor.AI sp. z o.o. (KRS 0001190083), Wroc\l{}aw. He is concurrently affiliated with the Estonian Entrepreneurship University of Applied Sciences, which supplied no funding. Rankfor.AI sells AI-visibility analytics to commercial clients, and this corpus supports research in the market it sells into. The corpus was built on, and is hosted by, Rankfor.AI infrastructure. Readers should weigh the findings accordingly. The authors have no professional or academic relationship with any individual they would ask to be excluded as a reviewer.

\textbf{Data availability.} Two subsets are public and require no request. PersonaGen-15K, a 14,955-persona stratified subset, is at \url{https://huggingface.co/datasets/rankfor/PersonaGen-15K} under CC BY-NC-SA 4.0, a licence it inherits from two of the four upstream corpora. A separate 5,000-persona set generated by a different pipeline, PersonaGen-Enterprise, is at \url{https://huggingface.co/datasets/rankfor/PersonaGen-Enterprise} under CC BY 4.0.

The full 1,031,732-persona corpus is not released. It is held by the authors and shared on request for non-commercial research use, with the requester and purpose recorded; referees of this manuscript are given access on request. We state plainly what this costs: no persistent identifier exists for the full corpus, and a reader cannot independently verify the Section 5 statistics without contacting us. The corpus file carries \texttt{sha256 5b7c0461\ldots}, quoted so that a reader who does obtain it can confirm it is the file these statistics were computed from. The two public subsets are what this paper offers for unrestricted reuse. A separate 5,000-persona enterprise subset, PersonaGen-Enterprise, is released independently under CC BY 4.0. The four source datasets are public on HuggingFace (NVIDIA Nemotron-Personas-USA; BSC-LT m-Personas; Orange PersonasForSalesbot; Tencent PersonaHub). The supply-side citation-provenance dataset referenced in Sections 2.3 and 6 is deposited separately (Zenodo 10.5281/zenodo.20829524, CC BY 4.0) and is a distinct resource from this corpus.

\textbf{Code availability.} The construction pipeline is not released. Section 4 states the deduplication thresholds, the embedding model, the enrichment model and its decoding parameters, the batching and the quality-filter rate, so the procedure can be reimplemented; the scripts themselves are held with the corpus and are covered by the same access procedure.

\begin{center}\rule{0.5\linewidth}{0.5pt}\end{center}

\begin{appendices}
\section{Persona Schema}\label{appendix-a.-persona-schema}

\begin{verbatim}
{
  "original_uuid": "string",
  "industry": "string",
  "market_context": "string (B2C | B2B | B2B2C | B2G)",
  "original_description": "string",
  "details": {
    "name": "string",
    "gender": "string",
    "role": "string",
    "demographics": "string (free text; not parsed
                     into age/location/profession)",
    "decision_weights": "float",
    "goals": ["string"],
    "pains": ["string"],
    "story": "string",
    "informational_needs": ["string"],
    "top_uncovered_need": ["string"],
    "search_behaviour": {
      "primary_intent": "string (informational | commercial
                          | transactional | navigational)",
      "typical_queries": ["string"],
      "search_triggers": ["string"],
      "preferred_sources": ["string"]
    }
  }
}
\end{verbatim}

\end{appendices}

\section*{References}
\small
\noindent [1] P. Aggarwal, V. Murahari, T. Rajpurohit, A. Kalyan, K. Narasimhan, A. Deshpande. GEO: Generative Engine Optimization. In Proceedings of the 30th ACM SIGKDD Conference on Knowledge Discovery and Data Mining (KDD), 2024.\par\vspace{3pt}

\noindent [2] T. Ge, X. Chan, X. Wang, D. Yu, H. Mi, D. Yu. Scaling Synthetic Data Creation with 1,000,000,000 Personas (PersonaHub). arXiv:2406.20094, 2024.\par\vspace{3pt}

\noindent [3] D. \.Zatuchin. Who Owns the AI Recommendation? A Multi-Industry Empirical Map of Brand Category Ownership Across Large Language Models. arXiv:2606.23057, 2026; dataset on Zenodo, 10.5281/zenodo.20788142 (CC BY 4.0). (Evidence register: 3,750 responses, 50 brands, 5 industries, 3 LLMs; mean Gini 0.28, 95\% CI 0.16 to 0.41; competitive vacuums in 8.0\% of queries.)\par\vspace{3pt}

\noindent [4] A. Cooper. The Inmates Are Running the Asylum. Sams / Macmillan, 1999.\par\vspace{3pt}

\noindent [5] A. Revella. Buyer Personas: How to Gain Insight into Your Customer's Expectations, Align Your Marketing Strategies, and Win More Business. Wiley, 2015.\par\vspace{3pt}

\noindent [6] J. Pruitt, J. Grudin. Personas: practice and theory. In Proceedings of the 2003 Conference on Designing for User Experiences (DUX), 2003.\par\vspace{3pt}

\noindent [7] X. Zhang, H.-F. Brown, A. Shankar. Data-Driven Personas: Constructing Archetypal Users with Clickstreams and User Telemetry. In Proceedings of the 2016 CHI Conference on Human Factors in Computing Systems, pp. 5350–5359, 2016.\par\vspace{3pt}

\noindent [8] J. An, H. Kwak, J. Salminen, S. Jung, B. J. Jansen. Imaginary People Representing Real Numbers: Generating Personas from Online Social Media Data. ACM Transactions on the Web, 12(4), 2018. DOI 10.1145/3265986.\par\vspace{3pt}

\noindent [9] J. Salminen, K. Guan, S. Jung, B. J. Jansen. A Survey of 15 Years of Data-Driven Persona Development. International Journal of Human-Computer Interaction, 37(18):1685–1708, 2021. DOI 10.1080/10447318.2021.1908670.\par\vspace{3pt}

\noindent [10] L. Long, et al. On LLMs-Driven Synthetic Data Generation, Curation, and Evaluation: A Survey. In Findings of the Association for Computational Linguistics (ACL), 2024.\par\vspace{3pt}

\noindent [11] L. P. Argyle, E. C. Busby, N. Fulda, J. R. Gubler, C. Rytting, D. Wingate. Out of One, Many: Using Language Models to Simulate Human Samples. Political Analysis, 31(3):337–351, 2023.\par\vspace{3pt}

\noindent [12] G. V. Aher, R. I. Arriaga, A. T. Kalai. Using Large Language Models to Simulate Multiple Humans and Replicate Human Subject Studies. In Proceedings of ICML, 2023.\par\vspace{3pt}

\noindent [13] A. Broder. A taxonomy of web search. ACM SIGIR Forum, 36(2):3–10, 2002.\par\vspace{3pt}

\noindent [14] B. J. Jansen, S. Y. Rieh. The seventeen theoretical constructs of information searching and information retrieval. Journal of the American Society for Information Science and Technology, 61(8):1517–1534, 2010.\par\vspace{3pt}

\noindent [15] D. \.Zatuchin. How Large Language Models Source Brand Reputation Across Languages and Markets. arXiv:2606.25787, 2026; dataset on Zenodo, 10.5281/zenodo.20829524 (CC BY 4.0). (128 brands, 12 markets, 167,551 URL-grounded citations; 85.7\% third-party.)\par\vspace{3pt}

\noindent [16] M. Sclar, Y. Choi, Y. Tsvetkov, A. Suhr. Quantifying Language Models' Sensitivity to Spurious Features in Prompt Design, or: How I Learned to Start Worrying about Prompt Formatting. In Proceedings of the Twelfth International Conference on Learning Representations (ICLR), 2024. arXiv:2310.11324.\par\vspace{3pt}

\noindent [17] M. Mizrahi, G. Kaplan, D. Malkin, R. Dror, D. Shahaf, G. Stanovsky. State of What Art? A Call for Multi-Prompt LLM Evaluation. Transactions of the Association for Computational Linguistics, 2024. arXiv:2401.00595.\par\vspace{3pt}

\noindent [18] N. Kunievsky, J. A. Evans. Measuring Intent Comprehension in LLMs. arXiv:2506.16584, 2025.\par\vspace{3pt}

\noindent [19] B. Atil, S. Aykent, A. Chittams, L. Fu, R. J. Passonneau, E. Radcliffe, G. R. Rajagopal, A. Sloan, T. Tudrej, F. Ture, Z. Wu, L. Xu, B. Baldwin. Non-Determinism of ``Deterministic'' LLM Settings. arXiv:2408.04667, 2024.\par\vspace{3pt}

\noindent [20] D. \.Zatuchin. Where Does the Noise Come From? A Variance-Components Decomposition of Non-Determinism in LLM Brand Answers. arXiv:2607.13304, 2026.\par\vspace{3pt}

\noindent [21] D. \.Zatuchin. The Language Blind Spot: How Query Language and Brand Recognition Tier Shape AI-Constructed Brand Reputation Across Twelve European Languages. arXiv:2606.23165, 2026; dataset on Zenodo, 10.5281/zenodo.20794390 (CC BY 4.0).\par\vspace{3pt}

\noindent [22] P. Laban, H. Hayashi, Y. Zhou, J. Neville. LLMs Get Lost In Multi-Turn Conversation. arXiv:2505.06120, 2025.\par\vspace{3pt}

\noindent [23] M. Miroyan, T.-H. Wu, L. King, T. Li, J. Pan, X. Hu, W.-L. Chiang, A. N. Angelopoulos, T. Darrell, N. Norouzi, J. E. Gonzalez. Search Arena: Analyzing Search-Augmented LLMs. In Proceedings of the International Conference on Learning Representations (ICLR), 2026. arXiv:2506.05334.\par\vspace{3pt}

\noindent [24] A. Chatterji, T. Cunningham, D. J. Deming, Z. Hitzig, C. Ong, C. Y. Shan, K. Wadman. How People Use ChatGPT. NBER Working Paper 34255, National Bureau of Economic Research, September 2025.\par\vspace{3pt}

\noindent [25] X. Chen, H. Wu, J. Bao, Z. Chen, Y. Liao, H. Huang. Role-Augmented Intent-Driven Generative Search Engine Optimization. arXiv:2508.11158, 2025.\par\vspace{3pt}

\noindent [26] W. Wang, W. Jiao, J. Huang, R. Dai, J.-t. Huang, Z. Tu, M. Lyu. Not All Countries Celebrate Thanksgiving: On the Cultural Dominance in Large Language Models. In Proceedings of the 62nd Annual Meeting of the Association for Computational Linguistics (ACL), pp. 6349--6384, 2024. DOI 10.18653/v1/2024.acl-long.345.\par\vspace{3pt}

\noindent [27] MatrAIx. MatrAIx\_Persona\_1M: a 999,847-row persona coreset with 1,290 categorical attributes. Hugging Face dataset, published 1 August 2026. \texttt{https://huggingface.co/datasets/MatrAIx2026/MatrAIx\_Persona\_1M}.\par\vspace{3pt}

\noindent ---\par\vspace{3pt}
\normalsize

\end{document}